\documentclass{iau} 
\usepackage{graphicx}

\def\gpy{{\rm ~Gpc}^{-3} {\rm ~yr}^{-1}}
\newcommand{\msun}{\ensuremath{\,M_\odot}}
\newcommand{\rsun}{\ensuremath{\,R_\odot}}
\newcommand{\zsun}{\ensuremath{\,Z_\odot}}
\newcommand{\yr}{\ensuremath{\,\rm yr}}
\newcommand{\msunyr}{\ensuremath{\,M_\odot \, \rm yr^{-1}}}
\newcommand{\ergs}{\ensuremath{\,\rm erg \, s^{-1}}}

\title[JD 11.~~HMXB as progenitors of GW sources] 
{High mass X-ray binaries \\ as progenitors of gravitational wave sources}

\author[Jakub Klencki \& Gijs Nelemans]   
{Jakub Klencki$^1$
 \and Gijs Nelemans$^1,^2$}

\affiliation{$^1$Department of Astrophysics/IMAPP, Radboud University, \\
P O Box 9010, NL-6500 GL Nijmegen, The Netherlands \\ email: {\tt j.klencki@astro.ru.nl, nelemans@astro.ru.nl} \\[\affilskip]
$^2$Institute of Astronomy, KU Leuven, \\ 
Celestijnenlaan 200D, B-3001 Leuven, Belgium
}

\pubyear{2018}
\volume{xxx}  
\jname{High Mass X-ray Binaries: illuminating the passage from massive binaries to merging compact objects}
\editors{A.C. Editor, B.D. Editor \& C.E. Editor, eds.}
\begin{document}

\maketitle

\begin{abstract}
X-ray binaries with black hole (BH) accretors and massive star donors at short 
orbital periods of a few days can evolve into close binary BH systems (BBH)
that merge within the Hubble time through stable mass transfer evolution.
From observational point of view, upon
the Roche-lobe overflow, such systems will most likely appear as ultra-luminous
X-ray sources (ULXs). To study this connection, we compute the mass transfer
phase in systems with BH accretors and massive star donors ($M > 15 \msun$)
at various orbital separations and metallicities using the MESA stellar evolution code. In the case of core-hydrogen
and core-helium burning donors (cases A and C of mass transfer)
we find the typical duration of super-Eddington mass transfer of up to $10^6$
and $10^5$ yr, with rates of $10^{-6}$ and $10^{-5} \msunyr$ , respectively. 
Given that roughly 0.5 ULXs are found per unit of star formation rate ($\msunyr$), 
and assuming that 10\% of all the observed ULXs form merging BBH, we estimate the rate
of BBH mergers from stable mass transfer evolution to be at most $10 \gpy$.

\end{abstract}

\firstsection 
\section{Introduction}

The first discovery of a gravitational wave signal from a binary black hole (BBH) merger
by the Advanced LIGO Interferometer in September 2015 \cite[(Abbott et al. 2016)]{Abbott16_firstBBH} revived the discussion on 
possible formation scenarios for double compact objects. A large number of channels have been put forth, 
especially in the case of BBH, 
including but not limited to the formation from isolated binaries through a common envelope event
\cite[(eg. Belczynski et al. 2016, Eldridge \& Stanway 2016, Klencki et al. 2018, Mapelli \& Giacobbo 2018, Kruckow et al. 2018)]
{B16Nat,Eldridge2016,Klencki2018,Mapelli2018,Kruckow2018}, or 
in a chemically homogenoues evolution regime \cite[(Mandel \& de Mink 2016, de Mink \& Mandel 2016, Marchant et al. 2016)]{Mandel2016,deMink2016,Marchant2016},
dynamical formation in globular clusters \cite[(eg. Rodriguez et al. 2016, Askar et al. 2017)]{Rodriguez2016a, Askar2017}
in nuclear clusters \cite[(Arca-Sedda \& Gualandris 2018)]{ArcaSedda2018}, or in disks of active galactic nuclei
\cite[(Antonini \& Rasio 2016, Stone et al. 2017)]{Antonini2016, Stone2017}, 
as well as formation channels involving triple stellar systems \cite[(Antonini et al. 2017)]{Antonini17}. 
Given the difficulty of distinguishing the formation channels based on the gravitational 
wave information alone (although there is some hope connected
to the measurement of the BH-BH spin-orbit misalignments, eg. \cite[Farr et al. 2017, 2018]{Farr2017,FarrHolzFarr2018}),
as well as a lack of electromagnetic counterpart to BBH detected so far, 
the contribution of various channels to the entire population of gravitational wave sources 
is usually estimated on theoretical grounds. 

In most theoretical scenarios, the estimates of the merger rate density of BBH fall within the range of about $\sim 0.1-20\gpy$.
An exception is the case of common envelope (CE) evolution channel where the merger rate 
could possibly be as high as $\sim 100-200\gpy$ \cite[(eg. Belczynski et al. 2016)]{B16Nat}. For comparison, the 
current observational limits on the merger rate density of BBHs inferred by the LIGO/Virgo 
are $13-212\gpy$ \cite[(Abbott et al. 2017)]{Abbott17_rates}. 

For the moment, the CE evolution channel is a promising candidate for the origin of at least some of the BBH 
detected by the LIGO/Virgo. However, our understanding of the CE phase itself is still very limited \cite[(Ivanova 2011)]{Ivanova2011}, 
and a number of known issues exist. In particular, recent studies of mass transfer stability from 
the giant donor stars \cite[(Woods \& Ivanova 2011, Pavlovkii et al. 2017)]{Woods2011,Pavlovkii2017} reveal that the mass transfer 
remains stable for a larger parameter space than previously thought, thus avoiding a CE evolution in the majority of cases. 
Additionally, very massive stars have been shown to stay rather compact throughout their evolution 
and potentially never reach the large radii and outer convective envelope layers that are at the heart of a CE formation 
channel for BBH mergers. 

Recently, \cite[van den Heuvel et al. (2017)]{vdHeuvel2017} pointed out the possibility of forming close BBH systems via 
stable and inconservative Roche-love overflow (RLOF) mass transfer from radiative giants onto stellar BHs in binaries with mass ratio
$q = M_{\rm donor} / M_{\rm accretor} \simeq 3.0-3.5$, as in the case of the SS433 system \cite[(Hillwig \& Gies 2008)]{Hillwig2008}.
They show that a few of the known Galactic double spectroscopic WR-O-type binaries \cite[(van der Hucht 2001)]{vdHucht2001}
may be progenitors of such an evolutionary path. Based on the number of such systems and the expected duration 
of the WR phase, van den Heuvel et al. (2017) estimate a galactic merger rate of close BBH systems produced by 
stable mass transfer as $\simeq 0.5 \times 10^{-5} \; \rm yr^{-1}$ [i.e. one merger every $2 \times 10^5$ years]. For comparison, 
assuming a star formation rate (SFR) of $\sim 1 \rm M_{\odot} yr^{-1}$ for the Milky Way, the CE channel typically 
produces one BBH merger every $5\times 10^6$ yr in the case of Solar metallicity ($\zsun = 0.02$) 
and every $1 \times 10^5$ yr for $Z = 0.1\zsun$ \cite[(Klencki et al. 2018)]{Klencki2018}. This rough estimate signifies that 
the channel proposed by van den Heuvel et al. could be a competitive source of BBH mergers, especially at $Z \sim \zsun$ metallicites, 
which is the dominant metallicity of massive star forming galaxies in the nearby Universe. 

The phase of a stable RLOF mass transfer in binaries with BH accretors and massive-star donors
will inevitably be observable as a luminous high-mass X-ray binary (HMXB) phase, 
and most likely even as an ultra-luminous X-ray source \cite[(ULXs, Rappaport et al. 2005)]{Rappaport2005}.
Here we will discuss the channel proposed by Van den Heuvel from the point 
of view of the connection between a population of BBH systems and the population of ULXs. 

\section{Forming close binary black holes through stable mass transfer}

\begin{figure}[b]
\begin{center}
 \includegraphics[width=0.8\columnwidth]{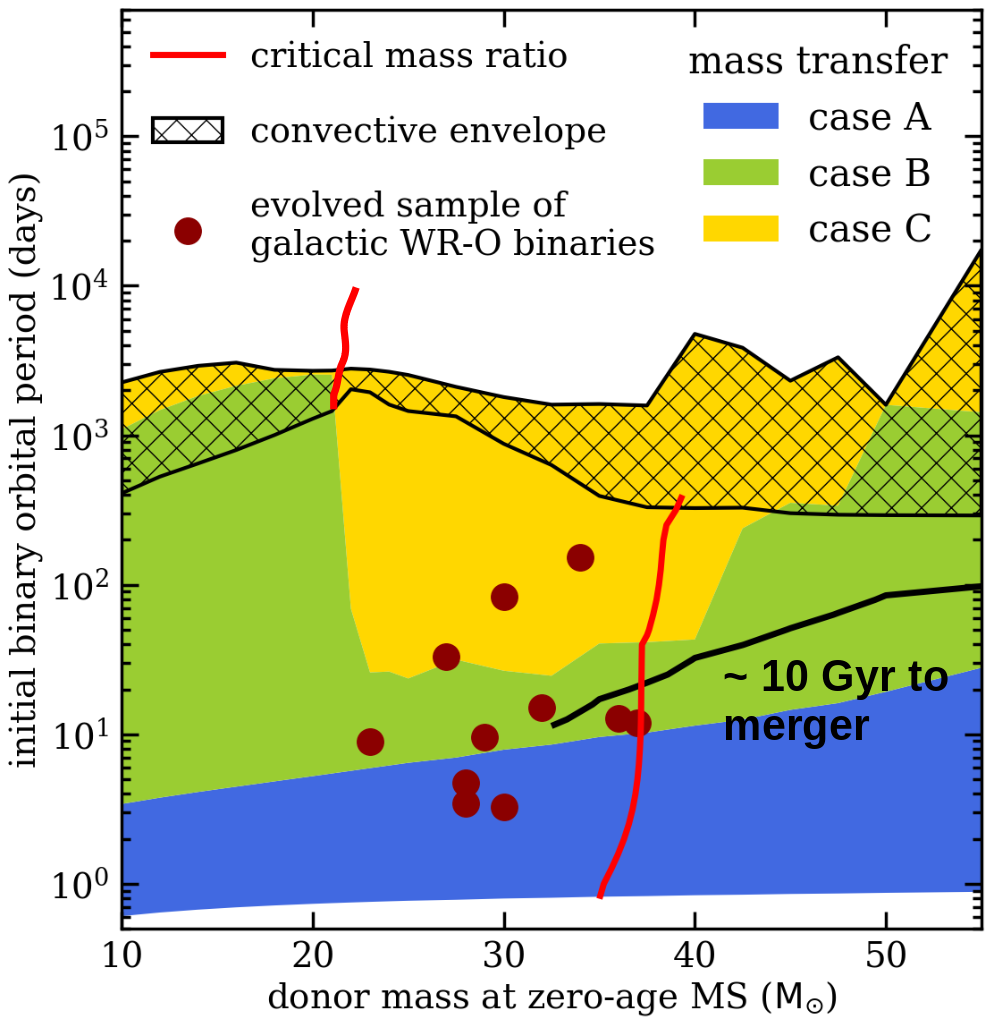} 
 \caption{Different cases of mass transfer from giant donors in a binary with a $10\msun$ black hole accretor
 at metallicity $0.0034$ (20\% Solar): case A, from a donor that is still a main-sequence star, case B, from a donor that is during the Hertzprung Gap
 evolution, and case C, from a donor that is already burning helium in the core. The mass transfer in systems to the left 
 of the red lines is expected to be stable. Dark red points indicate estimated parameters of BH-O-star systems that are likely descendants
 of a sample of known spectroscopic WR-O binaries (see text for the details). The black line indicates an estimated upper initial 
 orbital period limit on the formation of close binary black hole systems that merge in less than 10 Gyr.  
 Note that in Solar metallicity the parameter space for case C mass transfer is much smaller (see also de Mink et al. 2008)}
   \label{fig1}
\end{center}
\end{figure}

\underline{\textit{Van den Heuvel et al. scenario}}. In order to form close BBH systems through stable mass transfer
one requires RLOF to occur in BH binaries with giant donors at orbital 
periods of at most $\sim 10$ days and relatively high mass ratios $q = M_{\rm donor} / M_{\rm accretor} \sim 3$. 
\cite[(van den Heuvel et al. 2017)]{vdHeuvel2017}. 
Such systems are natural outcomes of a futher evolution of Wolf-Rayet-O-type star binaries, a few tens of which  
are known in the Milky Way and Magellanic Clouds \cite[(van der Hucht 2001)]{vdHucht2001}. The high values of $q$ 
make it possible for a stable mass transfer to reduce the binary orbital period enough (to values of the order of $\sim 1$ day), so 
that the system is going to merge due to gravitational wave emission within the Hubble time.
Alternatively, if the matter that is lost from the system during the inconservative mass transfer phase has high angular
momentum (i.e. higher than the usually assumed orbital angular momentum in the proximity of the accretor) then smaller mass ratios 
could also lead to close enough BBH systems.

{\underline{\it Stability of mass transfer from giant donors}}. In the first approximation, the stability of mass transfer in a binary depends mostly 
on the mass ratio of the components: $q = M_{\rm donor} / M_{\rm accretor}$. If the mass ratio is lower than some critical value, i.e. $q < q_{\rm crit}$
then the mass transfer will be stable, proceeding on a thermal or nuclear timescale of the donor star. If the mass ratio
is too high ($q > q_{\rm crit}$) then at some point the Roche lobe of the donor star shrinks more quickly with mass 
loss then the donor radius. This can cause dynamical instability, leading to a very high mass transfer rate and ultimately
to a CE evolution. In general, the value of $q_{\rm crit}$ depends on the exact structure of
the donor star. For example, the most immediate adiabatic response of the donor envelope to mass loss is sensitive to the entropy profile
\cite[(Ge et al. 2010, Ivanova 2015, Pavlovskii et al. 2017)]{Ge2010, Ivanova2015, Pavlovskii2017}. In a simplified picture, this can be summarized as follows:
in the case of donors with radiative envelope $q_{\rm crit} \approx 3.5$ \cite[(Ge et al. 2010)]{Ge2010}, 
whereas in the case of donors with deep enough layers of outer convective envelope
\footnote{eg. at least 10\% deep in mass coordinate} $q_{\rm crit} \approx 1.5-2.2$ \cite[(Pavlovskii \& Ivanova 2015)]{Pavlovkii2015}.

To investigate what types of donor envelopes expected possible in binaries with BH accretors depending on the orbital period,
we computed evolutionary models of single massive ($M > 15 \msun$)
stars at different metallicities using the MESA code \cite[(Paxton et al. 2011,2015)]{Paxton2011,Paxton2015}
until the end of core helium burning. While further 
evolution may result in additional mass transfer phases, they would not be long-lasting because there is not much time left until the core collapse 
at this point ($\sim 10^4$ yr). 
We model convection by using the mixing-length theory \cite[(B{\"o}hm-Vitense 1958)]{BohmVitense1958} with a mixing-length parameter $\alpha = 1.5$, and we adopt the Ledoux criteria 
for convection. We account for semi-convection following \cite[(Langer et al. 1983)]{Langer1983} with an efficiency parameter $\alpha_{\rm SC} = 1.0$. 
We also account for overshooting by applying a step overshooting formalism with an overshooting length of 0.385 pressure scale heights. 
Stellar winds are modeled in a way described in \cite[Brott et al. (2011)]{Brott2011}.

In Fig.~1 we show, as a function of the donor mass ($M_d$) and the initial binary orbital period ($P_{\rm init}$), different cases of a possible mass transfer
in a binary with a $10\msun$ BH accretor:
from a donor star that is still during its main sequence (MS) evolution (case A), from a donor that is crossing the Hertzprung Gap (HG) during 
hydrogen shell burning (case B), and from a donor that is already burning helium in the core (case C). The red lines mark critical mass ratios
whereas the black hatched area indicates donors with convective outer envelopes. Note that the vast majority of cases in Fig.~1 are donors with radiative envelopes. 
The chosen metallicity for this plot is $20\%$ the Solar metallicity, which is a common metallicity of many ULX-host galaxies \cite[(Mapelli et al. 2010)]{Mapelli2010}.
At Solar metallicity we find that the parameter space for a case C mass transfer is very small due to $15-30 \msun$ stars expanding all the way until the giant branch before the 
core-helium ignition (see also de Mink et al. 2008). 

Additionally, we show the estimated parameters of BH-O-star binaries that are expected to be the further evolutionary stage of a sample of double-spectroscopic 
WR-O binaries with well known component masses \cite[(van der Hucht 2001)]{vdHucht2001}. To do so, we make the same assumptions about the WR lifetime, mass-loss rate, and the BH formation
as \cite[van den Heuvel et al. (2017)]{vdHeuvel2017}, see their Sect.~3.4.

{\underline{\it Parameter space for merging BBH}}. For each pair $M_d$, $P_{\rm init}$ that falls into case B or case C mass transfer regimes in Fig.~1
we can estimate what would be the parameters of a resulting BBH system 
by making a few assumptions about the further evolution (again, similar to the assumptions made by Van den Heuvel et al. 2017).
In particular, we assume that the entire envelope mass of the donor is lost during the mass transfer (producing a BH-WR system),
that the mass transfer is 100\% non-conservative, and
that the matter is expelled from the proximity of the accreting BH having its orbital specific angular momentum.
We also account for WR winds and estimate a compact object formation (BH or NS) from the 'rapid' supernova engine prescription 
from Fryer et al. (2012). All BHs were assumed to form in direct collapse (i.e. no natal kick) with 10\% of the collapsing core mass lost in neutrinos.
Note that in some case when $M_d$ is too small a NS is formed instead of a secondary BH.
\\
The assumption that each donor loses its entire envelope mass can be justified 
and easily implemented in the case B and case C mass transfer systems for which 
most of the helium core has already be formed during the entire MS evolution, and the core-envelope boundary can be defined. 
Case A mass loss from a star, unless the RLOF occurs at the very end of MS, can significantly reduce the final helium core mass 
of the donor with respect to a single star evolution case. Case A region in Fig.~1 thus requires detailed binary evolution modelling. 
\\
For every BBH formed from case B and case C mass transfer systems we then calculate what would be the delay time between the formation of a BBH 
and the merger due to gravitational wave emission. With the black line in Fig.~1 we show a threshold delay time value of $\sim$ 10 Gyr; systems 
bellow the black line could merge within the Hubble time, while those above it would still be too wide after a stable mass transfer episode. 
The purpose of this rough estimate is to show, after \cite[van den Heuvel et al. (2017)]{vdHeuvel2017}, that it is possible to form BBH systems through stable mass 
transfer and a ULX phase with short enough orbital periods so that they will contribute to the BBH merger population. Notably, the larger the 
component mass ratio at RLOF the more likely it becomes that the final product will be a merging BBH. We wish to highlight at this point 
that the recent study of mass transfer stability of binaries with BH accretors and giant radiative donors \cite[(Pavlovskii et al.)]{Pavlovkii2017} suggests that
the critical mass ratio in their case can be larger than the $q_{\rm crit} = 3.5$ value plotted in Fig.~1, being even as high as $\sim6-8$ in some cases. 

\section{ULXs as progenitors of BBH mergers}

From an observational point of view, 
binaries of compact object and massive donors that transfer mass through an accretion disc 
are high luminosity HMXBs. 
The mass transfer rates in the case of RLOF in BH binaries are most likely
super-Eddington already from donors of a few Solar masses 
\cite[(Podsiadlowski et al. 2003, Rappaport et al. 2005)]{Podsiadlowski2003, Rappaport2005}, 
and even more so in the case of more massive donors $M > 15-20 \msun$.
Such systems are primary candidates for ultra-luminous X-ray (ULX) sources with $L_X > 10^{39} \; \rm erg/s$, 
both on the ground of theoretical models of accretion 
\cite[(Lipunova 1999, Poutanen et al. 2007, Lasota et al. 2016)]{Lipunova1999, Poutanen2007, Lasota2016} and
GRMHD simulations of super-critical accretion disks around BHs 
\cite[(Sadowski et al. 2014, McKinney et al. 2014, Sadowski \& Narayan 2016)]{Sadowski2014, McKinney2014, Sadowski2016}.
The rate of formation of BBH mergers through stable mass transfer is thus anchored to the number of ULXs systems observed. 
We can take advantage of this fact in order to put an upper limit on the local merger rate of BBHs from the Van den Heuvel scenario. 
If the number of ULXs per unit of SFR is $n_{\rm ULX} \, (\msun \yr^{-1})^{-1}$ and an average duration of the ULX phase is $t_{\rm ULX}$ 
then one ULX is formed per every $M_{\rm SF;ULX} = t_{\rm ULX} / n_{\rm ULX} \;(\msun)$ stellar mass formed. 
Assuming that a fraction $f_{\rm BBH}$ of ULXs will form close BBHs, and that the delay time distribution of this
BBH population is $dN/dt_{\rm del}$, the local 
merger rate of BBHs $R_{\rm BBH;loc} \, (\yr^{-1})$ that formed from ULXs can be expressed as:
\begin{equation}
 R_{\rm BBH;loc} = \int_{t_{\rm del} = 0}^{t_{\rm Hubble}} {\rm SFR}(z(t_{\rm del}))\times\frac{n_{\rm ULX}}{t_{\rm ULX}}\times f_{\rm BBH} \times \frac{dN}{dt_{\rm del}} dt_{\rm del}
\end{equation}
where $\rm SFR(z) \, \msunyr$ is the cosmic star formation rate, and $z(t_{\rm del})$ is the redshift corrsponding 
to a lookback time equal $t_{\rm del}$. The values of $n_{\rm ULX}$ and $\rm SFR(z)$ are determined observationally.
We will now take a look at other terms in the above formula. 

The delay time distribution of BBH mergers $dN/dt_{\rm del}$
in general depends on a particular formation scenario. 
However, if the distribution of the semi-major axes of the newly
formed BH-BH binaries can be described by a power law $dN/da \approx a^{-\beta}$, 
then the distribution of the delay times is $dN/dt_{\rm del} \approx t_{\rm del}^{-\alpha}$
where $\alpha = (3+\beta)/4$ because the delay time is proportional to $t_{\rm del} \propto a^4$.
One can see that even if $\beta$ varies in an extreme range from 0 to 7 then $\alpha$ is between 0.75 and 2.
Thus, for most astrophysical scenarios, $\alpha$ is constrained to $1 < \alpha < 2$.

\begin{figure}[b]
\begin{center}
 \includegraphics[width=0.98\columnwidth]{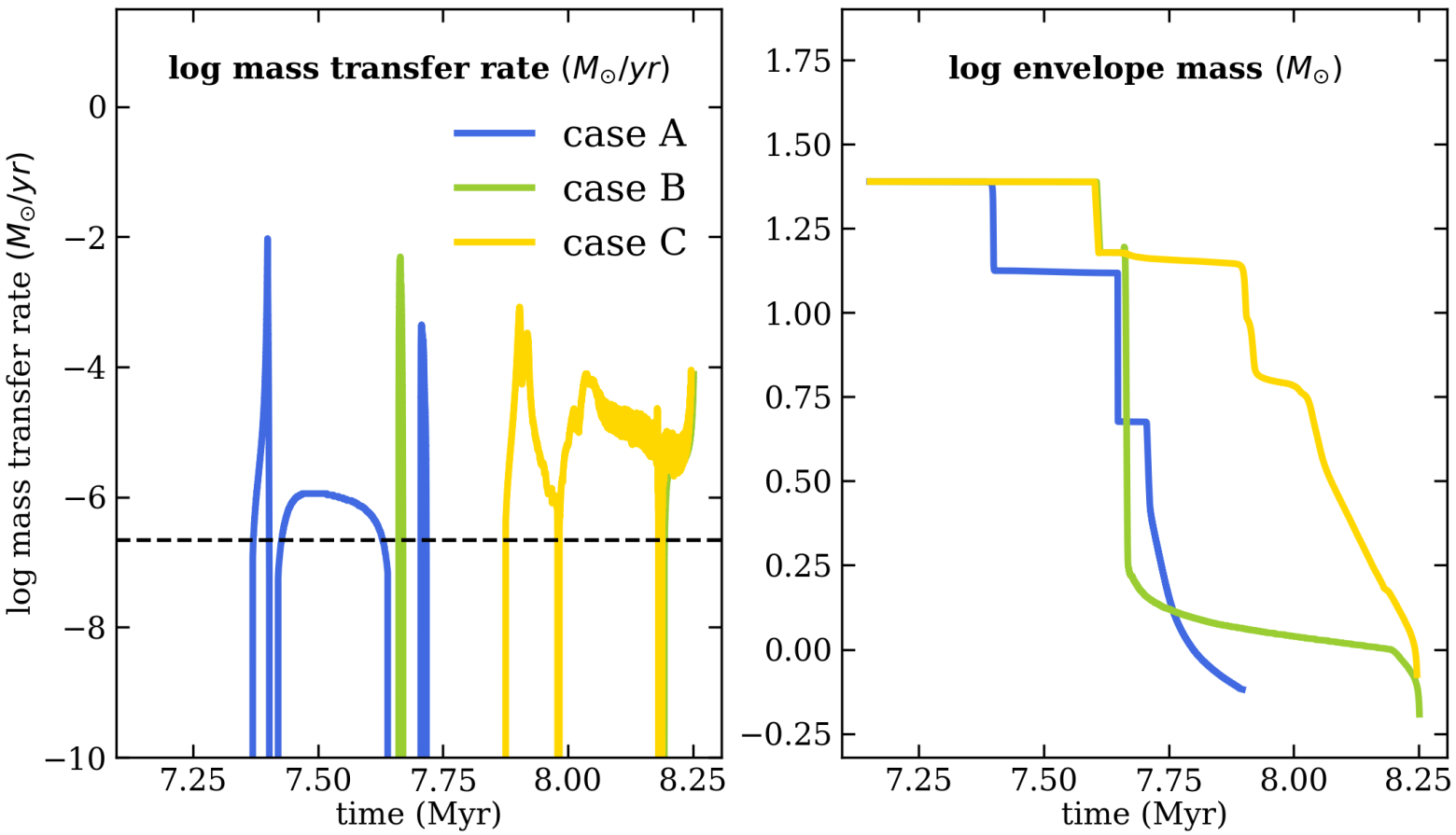} 
 \caption{Examples of time-evolution of the mass transfer rate (left) and the donor envelope mass (right) 
 computed with MESA for binaries of $10\msun$ BH accretors and $25\msun$ giant donors at metallicity $0.0034$ (20\% Solar).
 Different colors indicate different 
 initial binary separations ($40$, $50$ and $150\rsun$, corresponding to orbital 
 periods of $\sim$ 5, 7, and 36 days, respectively), and hence different cases of mass transfer: case A, case B, 
 and case C (see also Fig.~1). The horizontal dashed line in the left panel indicates Eddington mass transfer rate. 
 Note that in case A and case C mass transfer the super-Eddington mass transfer continues for a duration of
 a few $\times 10^5 \yr$, which is significantly longer than thermal timescale of the donor star. 
 We find a significant contribution of the case C mass transfer only in subsolar metallicity ($\lesssim 0.2 \zsun$).}
   \label{fig2}
\end{center}
\end{figure}

It is beyond the scope of this paper to estimate the fraction of ULXs that produce merging BBH: $f_{\rm BBH}$.
Hence, we treat it as a free parameter. In Fig.~1 one can see that only systems with large enough mass ratios 
and within the optimal orbital period range can end their evolution as close enough BBH systems. In particular, 
except a few promising candidates, most of the observed WR-O binaries are not expected to fall in the right parameter space. 
Additionally, it is known that a fraction of ULXs host NS accretors \cite[(Kaaret et al. 2017)]{Kaaret17}. BH X-ray binaries 
with donor masses of a few Solar masses can also contribute to the ULX population 
\cite[(Podsiadlowski et al. 2003, Rappaport et al. 2005)]{Podsiadlowski2003, Rappaport2005}, 
most likely the low-luminosity end of up to a few $\times 10^{39} \ergs$.
For all these reason we consider a fraction $f_{\rm BBH} = 0.1$ to be a conservative upper limit. 

In X-ray binaries with stellar BH accretors, the average duration of ULX phase $t_{\rm ULX}$ is the duration of super-Eddington (or close to such) mass transfer.
The Van den Heuvel scenario requires substantially large mass ratios $q \gtrsim 3$ at RLOF so that the orbital separation can shrink sufficiently. 
For such $q$ values the mass transfer will be launched on the thermal timescale of the donor star (the Kelvin-Helmholz timescale, $t_{\rm KH}$).
In the case of massive stars $t_{\rm KH} \approx 10^4$ yr or less. Naivly, one could take this value for the duration of the entire 
mass transfer episode (eg. as in discussion of Mineo et al. 2012). 

In practice, the issue is more complicated. In addition to the component mass ratio, the mass transfer rate (and hence its duration)
also depend on the response of the donor star to mass loss, which in turn depends on the donor structure in each particular case. 
Moreover, the component mass ratio will be changing over the course of the entire mass transfer, and even though the mass transfer 
may be proceeding on a thermal timescale in the beginning, it is possible that once the $q$ becomes lower the mass transfer rate will slow down 
to a nuclear timescale. 

In order constrain the average $t_{\rm ULX}$ more accurately, we have run a grid of binary evolution simulation models 
using the MESA code (with the same assumptions as described before). We have investigated 
cases of binaries with BH accretors ($M_{\rm BH} = 10$ or $20\msun$) and massive donors ($M_d$ from $15$ to $30\msun$ in $2.5\msun$ steps)
for different initial binary separations (18 different values from $a = 20\rsun$ to $2100\rsun$).

A systematic study of the computed evolutionary tracks of BH binaries across different metallicites in underway (Klencki et al. in prep). 
Here, in Fig.~2 we present a representative example of how typically the mass transfer proceeds, depending mostly on the 
evolutionary stage of the donor star (i.e. the mass transfer case; see also Fig.~1).
We plot three different instances of a time-evolution of the mass transfer rate and the donor envelope mass in 
binaries with $10\msun$ BH accretors and $25\msun$ donor stars at three different initial separations: $40$, $50$ and $150\rsun$,
which correspond to case A, case B, and case C mass transfer types respectively. Note that in all the three cases 
the donor has a radiative envelope at the point of mass transfer. 

One can see that in all the three cases the initial mass transfer phase is very rapid, proceeding approximately on a thermal timescale 
of the donor star: mass transfer rates of the order of $10^{-3}\div10^{-2} \msunyr$, for a duration of $0.5\div2 \times 10^4 \yr$.
This is the result of a relatively high mass ratio $q \approx 2.5$ at the moment of RLOF. 

In cases A and C the mass transfer slows down when the component mass ratio drops to about $q \sim 1.4$.
Afterwards, there is enough mass left in the donor envelope to power another longer phase of nuclear timescale mass transfer that is also super-Eddington.
Because the nuclear timescale of helium core burning is $\sim 10$ times shorter than that of hydrogen core burning, the nuclear mass transfer rate
in case C is about $\sim 10$ times higher: $10^{-5} \msunyr$ in case C compared to $10^{-6} \msunyr$ in case A. 
In case B mass transfer, on the other hand, 
this rapid phase continues for as long as the core collapses (roughly until the core helium ignition) and causes
the propagation of the hydrogen burning shell through the star that is pushing the outer stellar layers even more outwards. Once this 
phase is completed there is so little mass left in the envelope that no further mass transfer occurs. Only in some cases of a very late case B 
mass transfer is this different, and similar to a case C evolution. 
Notably, in case A and case C the secondary mass transfer episodes can last for a few $\times 10^5 \yr$ up to a few $\times 10^6 \yr$
if the RLOF occurs early during the MS evolution, which is much longer than the initial rapid thermal mass transfer phase. In the 
case of super-critical accretion discs the X-ray luminosity is a slow logarithmic function of the mass transfer rate $L_X \propto L_{\rm Edd} (1 + \alpha \, \dot{m})$ 
where $\dot{m} = \dot{M} / \dot{M_{\rm Edd}}$ and $\alpha$ is of the order of unity (eg. $\alpha = 3/5$ in an advection dominated 
disk with winds, Poutanen et al. 2007). For that reason, even though the mass transfer rates are significantly higher during 
the short-lived thermal mass transfer phases, we are more likely to observe ULXs during the longer-lasting
mass transfer episodes with $\dot{M}$ of the order of $10^{-6}\div10^{-5} \msunyr$. Especially if the radiation is beamed 
and the beaming is proportional to $\propto \dot{m}^2$ as suggested by \cite[King et al.]{King2016} (2016; although it is possible that
the beaming is limited to a factor of a few due to advection, see \cite[Lasota et al. 2016]{Lasota2016} and also GRMHD simulations
of \cite[Sadowski \& Narayan 2016]{Sadowski2016}).

\section{Upper limit on BBH merger rate from stable mass transfer}

Following the reasoning from the begining of Section 3, we can find an upper limit on the merger rate of BBH systems that formed through 
stable mass transfer from the observed number of ULXs. 
ULX-oriented surveys have shown that, by averaging over different galaxy types and different metallicites, there is roughly 0.5 ULX observed 
per unit of star formation rate ($\msunyr$, \cite[Mapelli et al. 2010, Swartz et al. 2011, Walton et al. 2011]{Mapelli2010,Swartz11,Walton11}).
\footnote{In fact, the X-ray luminosity function of ULX sources
appears to be a natural continuation of the X-ray luminosity function of X-ray binaries, Mineo et al. 2012.}
For an average duration of the ULX phase $T_{\rm ULX} (\yr)$, this implies that 1 ULX source is formed per every $0.5 \times T_{\rm ULX} (\msun)$ 
stellar mass formed. Assuming that a fraction $f_{\rm BBH}$ of ULXs are progenitors of BBH systems that are going to merge within the Hubble time, 
and that a typical delay time from the formation of BBH until the merger is $t_{\rm del} = 1$ Gyr, we can integrate the formation rate of ULXs 
over the cosmic SFR \cite[(taken from Madau \& Dickinson 2014)]{Madau2014}
to calculate the local (i.e. redshift z = 0) merger rate density of BBH formed from ULXs (equation 3.1):
\begin{equation}
 R_{\rm BBH;loc} \approx 1 \times \frac{f_{\rm BBH}}{0.1} \times \frac{10^{6} \yr}{T_{\rm ULX}} \; \gpy
\end{equation}
On the basis of our simulations of mass transfer in BH binaries,
we expect the average duration of the ULX phase in BH HMXBs to be of the order of $T_{\rm ULX} \approx $ a few $\times 10^5 \yr$. Assuming 
that $f_{\rm BBH} = 0.1$ of all the ULXs form BBH
systems that merge within the Hubble time, we obtain an upper limit on the local merger rate $R_{\rm BBH;loc} \lesssim 10 \gpy$.
We note that this result does not depend significantly on the choice of a particular delay time distribution.

\section{Summary}

We have studied the mass transfer in binaries comprised of stellar BH accretors
and massive star donors ($M > 15\msun$) that overflow their Roche-lobes 
by computing single and binary evolution models with the MESA code. 
In the case of donors with radiative envelopes, which we find to be the ones that dominate the parameter space of possible 
systems, the mass transfer is stable up to mass ratios $q = M_{\rm donor} / M_{\rm accretor}$ of at least $q \sim3-3.5$.
Such systems are also the most likely descendants of a sample of known O-WR binaries (see Fig.~1). 
For $q \gtrsim 2$ at the point of RLOF, the mass transfer launches on a thermal timescale of the donor star.
In the case of massive star donors this implies mass transfer rates of $10^{-3}\div10^{-2} \msunyr$. 
It is typically assumed that during such a thermal mass transfer the entire donor envelope is lost, and 
that this entire phase lasts for only about $10^3\div10^4$ yr (eg. \cite[Mineo et al. 2012]{Mineo2012}). 
Our simulations confirm this simple picture in the case of Solar metallicity, at which the massive stars expand significantly 
on a thermal timescale during the Hertzprung gap evolution before igniting helium in the core. This expansion helps to sustain 
a significant mass loss rates of $10^{-3}\div10^{-2} \msunyr$ until only a small envelope ($\sim 1 \msun$) is left. However, 
in the case of lower metallicity stars (eg. $Z = 0.2\zsun$), thanks to steeper density gradients and higher temperatures in the stellar cores, 
the helium is ignited earlier in the evolution, and the thermal expansion towards the giant branch is slowed down to a nuclear 
timescale at a smaller radii (eg. $\sim 100-200 \rsun$) than in the case of Solar metallicity stars (eg. $\sim 1000 \rsun$). 
This allows for a mass transfer from slowly-expanding
core-helium burning giants (which we refer to as case C mass transfer) in a significant number of cases (see Fig.~1, and 
also \cite[de Mink et al. 2008]{deMink2008}). In such cases, the phase of a rapid thermal mass transfer lasts only until 
the component mass ratio drops to about $q = 1.4$. Afterwards, there is still enough mass left in the envelope to power 
a secondary longer-lasting mass transfer phase on a nuclear timescale (see Fig.~2). The typical duration of that phase 
is of the order of a few $\times 10^5$ yr, with mass transfer rates of about $10^{-5}\div10^{-4} \msunyr$. 
In the case of a $10 \msun$ BH accretor, for which the Eddington mass transfer rate is roughly $2.2 \times 10^{-7} \msunyr$, 
all these mass transfer rates are super-Eddington. Such systems are thus expected to be observed as ultra-luminous X-ray sources 
(if viewed from the right angle given that the emission can be beamed).
Because in the case of super-critical accretion disks
the X-ray luminosity scales slowly with the mass transfer rate 
$\dot{M}$ as $L_{\rm} \propto L_{\rm Edd} [1 + ln(\dot{M}/\dot{M_{\rm Edd}})]$ (eg. \cite[Lipunova 1999, Poutanen et al. 2007]{Lipunova1999,Poutanen2007}), 
we argue that longer-lasting mass transfer phases with $\dot{M} \sim 10^{-5} \msunyr$ are more likely to observed than the shorter duration episodes 
of thermal mass transfer $\dot{M} \sim  10^{-3} \msunyr$. The case of less massive donors ($M < 15\msun$) was studied previously by  
\cite[Podsiadlowski et al. (2003) and Rappaport et al. (2005)]{Podsiadlowski2003, Rappaport2005}, who found $\dot{M}$ of up to $10^{-6} \msunyr$, and 
durations of at least several Myr. Consequently, the average duration of the ULX phase in the case of binaries with BH accretors 
is most likely of the order of $T_{\rm ULX} \approx 10^6$ yr. This agrees with the estimated ages of nebulae around some of the ULXs being  $\sim$ 1 Myr
\cite[(Pakull \& Mirioni 2002, Abolmasov \& Moiseev 2008)]{Pakull2002,Abolmasov2008}.

As an immediate application, we use the estimated value of $T_{\rm ULX}$ to discuss an upper limit on the merger rate of BBH systems 
that can form through stable mass transfer evolution, as suggested by van den Heuvel et al. (2017). Given that there is roughly 
0.5 ULXs observed per unit of star-formation rate, and assuming that 10\% of all the ULXs form merging BBH systems through the 
van den Heuvel scenario, we estimate the local merger rate of BBH formed this way to be about $R_{\rm BBH;loc} \approx 1\div10 \gpy$.

\begin{acknowledgements}
We thank Tomasz Bulik for a helpful discussion about the delay time distribution of double compact objects. 
The authors acknowledge  support  from  the  Netherlands  Organisation  for  Scientific  Research  (NWO).
\end{acknowledgements}


\begin{thebibliography}{}


\bibitem[{{Abbott} \etal\ (2016){Abbott}, {Abbott}, {Abbott}, {Abernathy},
  {Acernese}, {Ackley}, {Adams}, {Adams}, {Addesso}, {Adhikari}, \&
  et~al.}]{Abbott16_firstBBH}
{Abbott}, B.~P., {Abbott}, R., {Abbott}, T.~D., \etal\ 2016, \textit{Physical Review
  Letters}, 116, 061102
  
 \bibitem[{{Abbott} \etal\ (2017){Abbott}, {Abbott}, {Abbott}, {Acernese},
  {Ackley}, {Adams}, {Adams}, {Addesso}, {Adhikari}, {Adya}, \&
  et~al.}]{Abbott2017_rates}
{Abbott}, B.~P., {Abbott}, R., {Abbott}, T.~D., \etal\ 2017, \textit{Physical Review
  Letters}, 118, 221101
  
\bibitem[{{Abolmasov} \& {Moiseev}(2008)}]{Abolmasov2008}
{Abolmasov}, P. \& {Moiseev}, A.~V. 2008, \textit{Rev. Mexicana Astron. Astrofis.}, 44, 301

\bibitem[{{Antonini} \& {Rasio}(2016)}]{Antonini2016}
{Antonini}, F. \& {Rasio}, F.~A. 2016, \textit{ApJ}, 831, 187

\bibitem[{{Antonini} \etal\ (2017){Antonini}, {Toonen}, \&
  {Hamers}}]{Antonini17}
{Antonini}, F., {Toonen}, S., \& {Hamers}, A.~S. 2017, \textit{ApJ}, 841, 77

\bibitem[{{Arca-Sedda} \& {Gualandris}(2018)}]{ArcaSedda2018}
{Arca-Sedda}, M. \& {Gualandris}, A. 2018, \textit{MNRAS}, 477, 4423

\bibitem[{{Askar} \etal\ (2017){Askar}, {Szkudlarek}, {Gondek-Rosi{\'n}ska},
  {Giersz}, \& {Bulik}}]{Askar2017}
{Askar}, A., {Szkudlarek}, M., {Gondek-Rosi{\'n}ska}, D., {Giersz}, M., \&
  {Bulik}, T. 2017, \textit{MNRAS}, 464, L36
  
\bibitem[{{B{\"o}hm-Vitense}(1958)}]{BohmVitense1958}
{B{\"o}hm-Vitense}, E. 1958, \textit{ZAp}, 46, 108
  
\bibitem[{{Brott} \etal\ (2011){Brott}, {de Mink}, {Cantiello}, {Langer}, {de
  Koter}, {Evans}, {Hunter}, {Trundle}, \& {Vink}}]{Brott2011}
{Brott}, I., {de Mink}, S.~E., {Cantiello}, M., \etal\ 2011, \textit{A\&A}, 530, A115

\bibitem[{{de Mink} \& {Mandel}(2016)}]{deMink2016}
{de Mink}, S.~E. \& {Mandel}, I. 2016, \textit{MNRAS}, 460, 3545

\bibitem[{{de Mink} \etal\ (2008){de Mink}, {Pols}, \& {Yoon}}]{deMink2008}
{de Mink}, S.~E., {Pols}, O.~R., \& {Yoon}, S.-C. 2008, in American Institute
  of Physics Conference Series, Vol. 990, First Stars III, ed. B.~W. {O'Shea}
  \& A.~{Heger}, 230--232

\bibitem[{{Farr} \etal\ (2018){Farr}, {Holz}, \& {Farr}}]{FarrHolzFarr2018}
{Farr}, B., {Holz}, D.~E., \& {Farr}, W.~M. 2018, \textit{ApJL}, 854, L9

\bibitem[{{Farr} \etal\ (2017){Farr}, {Stevenson}, {Miller}, {Mandel}, {Farr},
  \& {Vecchio}}]{Farr2017}
{Farr}, W.~M., {Stevenson}, S., {Miller}, M.~C., {et~al.} 2017, \textit{Nature}, 548, 426

\bibitem[{{Ge} \etal\ (2010){Ge}, {Hjellming}, {Webbink}, {Chen}, \&
  {Han}}]{Ge2010}
{Ge}, H., {Hjellming}, M.~S., {Webbink}, R.~F., {Chen}, X., \& {Han}, Z. 2010,
  \textit{ApJ}, 717, 724
  
\bibitem[{{Hillwig} \& {Gies}(2008)}]{Hillwig2008}
{Hillwig}, T.~C. \& {Gies}, D.~R. 2008, \textit{ApJL}, 676, L37
  
\bibitem[{{Ivanova}(2015)}]{Ivanova2015}
{Ivanova}, N. 2015, {Binary Evolution: Roche Lobe Overflow and Blue
  Stragglers}, ed. H.~M.~J. {Boffin}, G.~{Carraro}, \& G.~{Beccari}, 179
  
\bibitem[{{Ivanova}(2011)}]{Ivanova2011}
{Ivanova}, N. 2011, in Astronomical Society of the Pacific Conference Series,
  Vol. 447, Evolution of Compact Binaries, ed. L.~{Schmidtobreick}, M.~R.
  {Schreiber}, \& C.~{Tappert}, 91
  
\bibitem[Kaaret \etal\ (2017)]{Kaaret17}
{Kaaret, P., Feng, H., \& Roberts T.~P.} 2017,
\textit{ARAA}, 55, 303
  
\bibitem[{{King} \& {Lasota}(2016)}]{King2016}
{King}, A. \& {Lasota}, J.-P. 2016, \textit{MNRAS}, 458, L10

\bibitem[{{Klencki} \etal\ (2018){Klencki}, {Moe}, {Gladysz}, {Chruslinska},
  {Holz}, \& {Belczynski}}]{Klencki2018}
{Klencki}, J., {Moe}, M., {Gladysz}, W., {et~al.} 2018, \textit{A\&A}, 619, A77

\bibitem[{{Langer} \etal\ (1983){Langer}, {Fricke}, \&
  {Sugimoto}}]{Langer1983}
{Langer}, N., {Fricke}, K.~J., \& {Sugimoto}, D. 1983, \textit{A\&A}, 126, 207

\bibitem[{{Lasota} \etal\ (2016){Lasota}, {Vieira}, {Sadowski}, {Narayan}, \&
  {Abramowicz}}]{Lasota2016}
{Lasota}, J.-P., {Vieira}, R.~S.~S., {Sadowski}, A., {Narayan}, R., \&
  {Abramowicz}, M.~A. 2016, \textit{A\&A}, 587, A13
  
\bibitem[{{Lipunova}(1999)}]{Lipunova1999}
{Lipunova}, G.~V. 1999, Astronomy Letters, 25, 508

\bibitem[{{Madau} \& {Dickinson}(2014)}]{Madau2014}
{Madau}, P. \& {Dickinson}, M. 2014, \textit{ARAA}, 52, 415

\bibitem[{{Mandel} \& {de Mink}(2016)}]{Mandel2016}
{Mandel}, I. \& {de Mink}, S.~E. 2016, \textit{MNRAS}, 458, 2634

\bibitem[{{Mapelli} \etal\ (2010){Mapelli}, {Ripamonti}, {Zampieri}, {Colpi},
  \& {Bressan}}]{Mapelli2010}
{Mapelli}, M., {Ripamonti}, E., {Zampieri}, L., {Colpi}, M., \& {Bressan}, A.
  2010, \textit{MNRAS}, 408, 234

\bibitem[{{Marchant} \etal\ (2016){Marchant}, {Langer}, {Podsiadlowski},
  {Tauris}, \& {Moriya}}]{Marchant2016}
{Marchant}, P., {Langer}, N., {Podsiadlowski}, P., {Tauris}, T.~M., \&
  {Moriya}, T.~J. 2016, \textit{A\&A}, 588, A50
  
\bibitem[{{McKinney} \etal\ (2014){McKinney}, {Tchekhovskoy}, {Sadowski}, \&
  {Narayan}}]{McKinney2014}
{McKinney}, J.~C., {Tchekhovskoy}, A., {Sadowski}, A., \& {Narayan}, R. 2014,
  \textit{MNRAS}, 441, 3177
  
\bibitem[{{Mineo} \etal\ (2012){Mineo}, {Gilfanov}, \& {Sunyaev}}]{Mineo2012}
{Mineo}, S., {Gilfanov}, M., \& {Sunyaev}, R. 2012, \textit{MNRAS}, 419, 2095

\bibitem[{{Pakull} \& {Mirioni}(2002)}]{Pakull2002}
{Pakull}, M.~W. \& {Mirioni}, L. 2002, ArXiv Astrophysics e-prints
  [\textit{astro-ph/0202488}]

\bibitem[{{Pavlovskii} \& {Ivanova}(2015)}]{Pavlovskii2015}
{Pavlovskii}, K. \& {Ivanova}, N. 2015, \textit{MNRAS}, 449, 4415

\bibitem[{{Pavlovskii} \etal\ (2017){Pavlovskii}, {Ivanova}, {Belczynski}, \&
  {Van}}]{Pavlovskii2017}
{Pavlovskii}, K., {Ivanova}, N., {Belczynski}, K., \& {Van}, K.~X. 2017,
  \textit{MNRAS}, 465, 2092
  
  \bibitem[{{Paxton} \etal\ (2011){Paxton}, {Bildsten}, {Dotter}, {Herwig},
  {Lesaffre}, \& {Timmes}}]{Paxton2011}
{Paxton}, B., {Bildsten}, L., {Dotter}, A., \etal\ 2011, \textit{ApJS}, 192, 3

\bibitem[{{Paxton} \etal\ (2015){Paxton}, {Marchant}, {Schwab}, {Bauer},
  {Bildsten}, {Cantiello}, {Dessart}, {Farmer}, {Hu}, {Langer}, {Townsend},
  {Townsley}, \& {Timmes}}]{Paxton2015}
{Paxton}, B., {Marchant}, P., {Schwab}, J., \etal\ 2015, \textit{ApJS}, 220, 15

\bibitem[{{Podsiadlowski} \etal\ (2003){Podsiadlowski}, {Rappaport}, \&
  {Han}}]{Podsiadlowski2003}
{Podsiadlowski}, P., {Rappaport}, S., \& {Han}, Z. 2003, \textit{MNRAS}, 341, 385

\bibitem[{{Poutanen} \etal\ (2007){Poutanen}, {Lipunova}, {Fabrika},
  {Butkevich}, \& {Abolmasov}}]{Poutanen2007}
{Poutanen}, J., {Lipunova}, G., {Fabrika}, S., {Butkevich}, A.~G., \&
  {Abolmasov}, P. 2007, \textit{MNRAS}, 377, 1187

\bibitem[{{Rappaport} \etal\ (2005){Rappaport}, {Podsiadlowski}, \&
  {Pfahl}}]{Rappaport2005}
{Rappaport}, S.~A., {Podsiadlowski}, P., \& {Pfahl}, E. 2005, \textit{MNRAS}, 356, 401
  
\bibitem[{{Rodriguez} \etal\ (2016){Rodriguez}, {Haster}, {Chatterjee},
  {Kalogera}, \& {Rasio}}]{Rodriguez2016a}
{Rodriguez}, C.~L., {Haster}, C.-J., {Chatterjee}, S., {Kalogera}, V., \&
  {Rasio}, F.~A. 2016, \textit{ApJL}, 824, L8
  
\bibitem[{{S{\c a}dowski} \& {Narayan}(2016)}]{Sadowski2016}
{S{\c a}dowski}, A. \& {Narayan}, R. 2016, \textit{MNRAS}, 456, 3929


\bibitem[{{S{\c a}dowski} \etal\ (2014){S{\c a}dowski}, {Narayan}, {McKinney},
  \& {Tchekhovskoy}}]{Sadowski2014}
{S{\c a}dowski}, A., {Narayan}, R., {McKinney}, J.~C., \& {Tchekhovskoy}, A.
  2014, \textit{MNRAS}, 439, 503

\bibitem[{{Stone} \etal\ (2017){Stone}, {Metzger}, \& {Haiman}}]{Stone2017}
{Stone}, N.~C., {Metzger}, B.~D., \& {Haiman}, Z. 2017, \textit{MNRAS}, 464, 946

\bibitem[Swartz \etal\ (2011)]{Swartz11}
{Swartz, D.~A., Soria, R., Tennant, A.~F., \& Yukita M.} 2011,
\textit{ApJ}, 741, 49 

\bibitem[{{van den Heuvel} \etal\ (2017){van den Heuvel}, {Portegies Zwart},
  \& {de Mink}}]{vdHeuvel2017}
{van den Heuvel}, E.~P.~J., {Portegies Zwart}, S.~F., \& {de Mink}, S.~E. 2017,
  \textit{MNRAS}, 471, 4256

\bibitem[{{van der Hucht}(2001)}]{vdHucht2001}
{van der Hucht}, K.~A. 2001, VizieR Online Data Catalog, 3215

\bibitem[Walton \etal\ (2011)]{Walton11}
{Walton, D.~J., Roberts, T.~P., Mateos, S., \& Heard V.} 2011,
\textit{MNRAS}, 416, 1844

\bibitem[{{Woods} \& {Ivanova}(2011)}]{Woods2011}
{Woods}, T.~E. \& {Ivanova}, N. 2011, \textit{ApJL}, 739, L48


%
%
%
%
%
%
%
%
%
%
%
%
%
%
%
%
%

\end{thebibliography}
\end{document}